\documentclass[twocolumn,prl,showpacs,superscriptaddress,preprintnumbers,amsmath,amssymb]{revtex4}

\usepackage{graphicx}
\usepackage{dcolumn}
\usepackage{bm}
\usepackage[]{color}

\begin{document}

\title{Correlation between ground state and orbital anisotropy in heavy fermion materials}

\author{T.~Willers}
  \affiliation{Institute of Physics II, University of Cologne, Z{\"u}lpicher Stra{\ss}e 77, 50937 Cologne, Germany}
\author{F.~Strigari}
  \affiliation{Institute of Physics II, University of Cologne, Z{\"u}lpicher Stra{\ss}e 77, 50937 Cologne, Germany}
\author{Z.~Hu}
 \affiliation  {Max Planck Institute for Chemical Physics of Solids, N{\"o}thnizer Stra{\ss}e 40, 01187 Dresden, Germany}
\author{V.~Sessi}
  \affiliation{European Synchrotron Radiation Facility (ESRF), B.P. 220, 38043 Grenoble C\'edex, France}
\author{N.~B.~Brookes}
  \affiliation{European Synchrotron Radiation Facility (ESRF), B.P. 220, 38043 Grenoble C\'edex, France}  
\author{E.~D.~Bauer}
	\affiliation{Los Alamos National Laboratory, Los Alamos, New Mexico 87545, USA}
\author{J.~L.~Sarrao}
	\affiliation{Los Alamos National Laboratory, Los Alamos, New Mexico 87545, USA} 
\author{J.~D.~Thompson} 
	\affiliation{Los Alamos National Laboratory, Los Alamos, New Mexico 87545, USA}  
\author{A.~Tanaka}
  \affiliation{Department of Quantum Matter, ADSM Hiroshima University, Higashi-Hiroshima 739-8530, Japan}
\author{S.~Wirth}
 \affiliation  {Max Planck Institute for Chemical Physics of Solids, N{\"o}thnizer Stra{\ss}e 40, 01187 Dresden, Germany}
\author{L.~H.~Tjeng}
 \affiliation  {Max Planck Institute for Chemical Physics of Solids, N{\"o}thnizer Stra{\ss}e 40, 01187 Dresden, Germany}
\author{A.~Severing}
  \affiliation{Institute of Physics II, University of Cologne, Z{\"u}lpicher Stra{\ss}e 77, 50937 Cologne, Germany}

\begin{abstract}
{The interplay of structural, orbital, charge and spin degrees of freedom is at the heart of many emergent phenomena, including superconductivity. Unraveling the underlying forces of such novel phases is a great challenge because it not only requires understanding each of these degrees of freedom, it also involves accounting for the interplay between them. Cerium-based heavy fermion compounds are an ideal playground for investigating these interdependencies and from our experiments we discover that there is a correlation between orbital anisotropy and the nature of the ground state in this material class. We measured the $4f$ crystal-electric field ground-state wave functions of the strongly correlated materials CeRh$_{1-x}$Ir$_x$In$_5$ with great accuracy using linear polarization-dependent soft x-ray absorption spectroscopy. These measurements establish that these wave functions determine the ground state properties of the substitution series, which covers long-range antiferromagnetic order, unconventional superconductivity, and coexistence of these two states. We discuss the impact of certain wave functions on magnetic order and anisotropic hybridization, as well as on the formation of a superconducting ground state. The latter may serve as a guide in the quest for enhancing superconducting transition temperatures, or even for new superconductors.}
\end{abstract}

\pacs{71.27.+a, 71.70.Ch, 74.70.Tx, 78.70.Dm}

\maketitle

Why do many chemically and structurally highly similar compounds develop different ground states? This seemingly simple question still eludes a straightforward description despite intense research. Yet, it is specifically pressing in view of the quest for a deeper insight into unconventional superconductivity.

We here investigate heavy fermion metals, i.e.\ rare earth or actinide materials, in which a plethora of phenomena including antiferromagnetism and superconductivity can be observed. In these compounds, the $f$~electrons hybridize with the conduction electrons ($cf$-hybridization) and, in analogy to the Kondo effect in diluted systems, the local magnetic moments can be screened in these so-called ``Kondo lattices'' at sufficiently low temperatures. However, the Kondo effect competes with the Ruderman-Kittel-Kasuya-Yosida (RKKY) interaction which typically favors long-range magnetic order. As a result of this competition a quantum phase transition from magnetically ordered to paramagnetic, more itinerant $f$ electron behavior can take place. The balance of both interactions depends on the exchange interaction $J_{ex}$ which can be tuned by external parameters such as pressure, magnetic field or doping \cite{Gegenwart_2008}. Non-Fermi liquid behaviour and, of interest here, unconventional superconductivity often occur in the vicinity of such quantum critical points. The quantum critical region of heavy fermion phase diagrams shows a striking similarity with cuprate (hole doping) and Fe pnictide (electron doping) phase diagrams\cite{Broun_2008,Norman_2011} -- all have in common an intriguing proximity of superconductivity and magnetism. Here heavy fermion materials may serve as prototypical systems because they can be made chemically very pure and the transitions of interest occur at very low temperatures where phonons are frozen out. In this respect they are the simpler and more straightforward correlated electron system to study.

\begin{figure}[t]
\centering
\includegraphics[width=0.45\columnwidth]{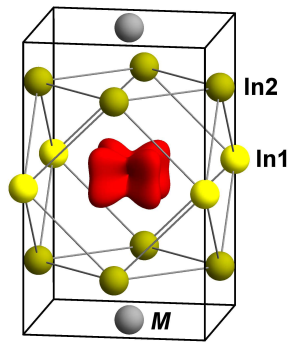}
\caption{Structure of Ce$M$In$_5$. Ce is represented by the angular distribution of the $4f$ CEF ground state orbital (red). The In1 (yellow), In2 (dark yellow) and the transition metal $M$ (gray) are labeled in the figure.}
\label{structure}
\end{figure}

\begin{figure}
\centering
\includegraphics[width=.9\columnwidth]{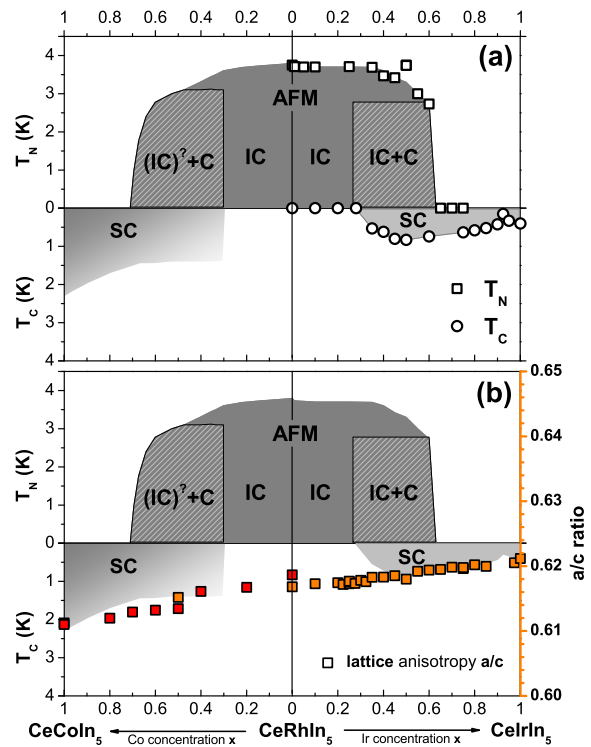}
\caption{(a) Phase diagram of CeRh$_{1-x}$Ir$_x$In$_5$ and CeRh$_{1-y}$Co$_y$In$_5$ as adapted from Refs.\,\cite{Pagliuso_2001,Pagliuso_2002a,Llobet_2005,Zapf_2001,Yokoyama_2008,Yokoyama,Ohira-Kawamura_2007,Ohira-Kawamura}. The N$\acute{e}$el temperature T$_N$ (white squares) and superconducting transition temperature T$_c$ (whites circles) are shown as function of the Ir and Co concentration. The incommensurate antiferromagnetic ordered phases (IC AF) and the commensurate ones (C~AF) are colored dark gray and hatched gray, respectively and the regions of superconductivity (SC) are marked light gray. The question mark on the Co side refers to the on-going discussion concerning the coexistence of IC and C AF order. The temperature scales of T$_N$ and T$_c$ are shown separately, one scale pointing up, the other one pointing down. (b) Lattice anisotropy a/c (orange squares) on the right scale for CeRh$_{1-x}$Ir$_x$In$_5$, CeRh$_{0.5}$Co$_{0.5}$In$_5$, and CeCoIn$_5$, plotted over the phase diagram. The lattice constants $a$ and $c$ are taken from Ref.\,\cite{Pagliuso_2001} (orange squares) and \cite{Zapf_2001} (red squares).}
\label{Fig2}
\end{figure}

The tetragonal heavy fermion family Ce\textit{M}In$_5$ with \textit{M}\,=\,Co, Ir, and Rh (HoCoGa$_5$-type structure, see Fig.\,\ref{structure}) is ideally suited for studying the interplay of antiferromagnetism, superconductivity and quantum criticality because their properties can be tuned easily by substitution \cite{Hegger_2000,Petrovic_2001,Petrovic_2001a,Bao_2000,Bao_2003,Park_2006,Thompson_2012,Aynajian_2012}. Figure~\ref{Fig2}a shows the substitution phase diagram of CeRhIn$_5$, where Rh is substituted by Ir or Co \cite{Pagliuso_2001,Pagliuso_2002a,Llobet_2005,Zapf_2001,Yokoyama_2008,Yokoyama,Ohira-Kawamura_2007,Ohira-Kawamura}.
For clarity we show the magnetically ordered and superconducting regions on separate scales (up and down, respectively). CeCoIn$_5$ and CeIrIn$_5$ are superconductors with transition temperatures of T$_c$\,=\,2.3 and 0.4\,K \cite{Petrovic_2001,Petrovic_2001a}.
CeRhIn$_5$, however, orders antiferromagnetically at T$_N$\,=\,3.8\,K. The magnetic order of CeRhIn$_5$ is incommensurate (IC AF) with the magnetic moments aligned in the $ab$ plane, propagating in a spiral along the tetragonal $c$-axis \cite{Bao_2000}. Substituting Rh with Ir or Co tunes the ground state away from magnetic ordering towards bulk superconductivity by passing through regions where magnetic order and superconductivity coexist, and where -- in the case of Ir substitution -- a second commensurate magnetic phase has been observed \cite{Llobet_2005}. A commensurate phase has also been observed on the Co side, but here the coexistence with the incommensurate order is still a matter of debate, possibly due to uncertainties of the samples' stoichiometry \cite{Yokoyama_2008,Yokoyama,Ohira-Kawamura_2007,Ohira-Kawamura}. We note that the Fermi surface of CeRhIn$_5$ resembles that of LaRhIn$_5$, i.e. the $4f$ electron of Ce remains localized and does not contribute to the Fermi surface volume, in contrast to the Ir and Co samples which show enlarged Fermi surfaces volumes, implying a more itinerant $f$ electron behaviour \cite{Haga_2001,Fujimori_2003,Harrison_2004,Shishido_2005,Settai2007,Allen2013}, especially in the regions of pure superconductivity \cite{PhysRevLett.101.056402,Shishido}.

On the search for interdependencies of physical parameters and ground state properties Bauer \textsl{et al.} speculated about a linear relationship between lattice anisotropy $c/a$ (here $a$ and $c$ are the lattice constants) and the superconducting transition temperature T$_c$ for the superconducting plutonium and cerium 115 compounds \cite{Bauer_2004}. We however, can see from Fig.\,\ref{Fig2}b that this relationship does not hold for almost half of the phase diagram. The yellow squares in Fig.\,\ref{Fig2}b show the the lattice anisotropy $c/a$ (or rather $a/c$) on top of the CeRh$_{1-y}$Co$_y$In$_5$ -- CeRh$_{1-x}$Ir$_x$In$_5$ phase diagram. The $a/c$ ratios are calculated from the values given in Ref.\,\cite{Pagliuso_2001} and in the following we will refer to $a/c$ only as lattice anisotropy. Figure~\ref{Fig2}b shows that Bauer \textit{et al.}'s scaling works rather well in the purely superconducting regions. However, for magnetically ordered, Rh-rich samples this linear dependency of T$_c$ on $a/c$ obviously breaks down. A closer inspection reveals that even in the region of phase coexistence there are significant deviations from a linear dependency (CeRh$_{1-x}$Ir$_x$In$_5$ with $0.3 \leq x \leq 0.45$). Consequently, the lattice anisotropy has no predictive power concerning the formation of a superconducting ground state, motivating us to look further for another parameter. In the following we concentrate on the CeRh$_{1-x}$Ir$_x$In$_5$ substitution series because the phase diagram is well defined (see Fig.~\ref{Fig2}a).

The importance of momentum-dependent hybridization and the impact of the anisotropic crystal-electric field (CEF) ground-state wave
functions has been discussed by several groups \cite{Pagliuso_2002,Thalmeier_2005,Burch_2007,Flint_2010,Haule_2010,Pourovskii_2014}. For the heavy fermion compound CeIrIn$_5$ Shim \textsl{et al.} have even made specific predictions on the basis of first principle DMFT calculations \cite{Shim_2007}. In particular, their calculations claim that the out-of-plane hybridization is stronger than the in-plane one, and that this is important to make the system more itinerant. Consequently the shape, i.e. the angular distribution, of the $4f$ orbital which is determined by the CEF should be relevant. This is demonstrated in Fig.\,\ref{structure} where the Ce~$4f$ orbital is drawn within the unit cell of Ce$M$In$_5$.  However, such an impact of the CEF on the ground state properties has never been proven experimentally, most likely due to the lack of accuracy of common methods. Here we present results of a systematic investigation of the CEF ground states and the different ground state properties of CeRh$_{1-x}$Ir$_x$In$_5$ using the soft x-ray absorption technique which specifically targets the $4f$ ground state wave function. The CeRh$_{1-y}$Co$_y$In$_5$ series appears less well suited due to unresolved questions with respect to a possible coexistence of IC and C magnetic order and the exact doping dependence of T$_c$ (cf. Fig.2a).

\section{Results}
\subsection{Soft x-ray absorption spectroscopy and choice of samples} We recently started using soft x-ray absorption spectroscopy (XAS) at the rare earth $M$-edges as a new method for measuring CEF ground state wave functions in heavy fermion compounds \cite{HansmannPRL100,WillersPRB81,StrigariPRB86}. Here the dipole selection rules for linearly polarized light provide the sensitivity to the ground state symmetry and allow its determination with unprecedented accuracy. The knowledge of excited states is not required for the ground state analysis as long as the data are taken at low enough temperatures so that only the ground state is populated, nor is data statistics or background hampering the result. This is a great advantage over more conventional methods like inelastic neutron scattering and single-crystal susceptibility. 

\begin{figure}[t]
\centering
\includegraphics[width=.8\columnwidth]{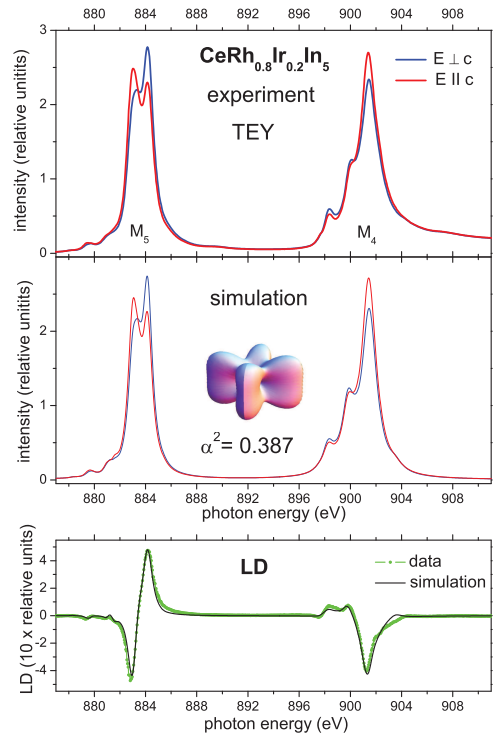}
\caption{Measured (top) and simulated (middle) linearly polarized XAS spectra of CeRh$_{0.8}$Ir$_{0.2}$In$_5$ at the $M_{4,5}$ edges at $T$\,=\,8\,K. The red spectra refer to measurements and simulations with E\,$\parallel$\,c and the blue ones with E\,$\bot$\,c. The inset in the middle panel shows the corresponding $4f$ spatial distribution. The bottom panel compares the measured (green dots) and simulated LD (black line) enlarged by a factor of ten.}
\label{Fig3}
\end{figure}

We have chosen the Ir concentrations $x$\,=\,0, 0.2, 0.3, 0.5, 0.75 and 1 of the CeRh$_{1-x}$Ir$_x$In$_5$ series for the liner polarized soft XAS experiment. Here $x$\,=\,0 and 0.2 cover the purely incommensurate antiferromagnetic region of the phase diagram with an almost identical T$_N$. The region of phase coexistence, which comprises an incommensurate, a commensurate and a superconducting state, is covered with the $x$\,=\,0.3 and 0.5 samples. We note that in both magnetic phases the magnetic moments are aligned antiferromagnetically in the basal plane and that the region of phase coexistence has been discussed in detail in Ref.\,\cite{Thompson_2012}. The magnetic order decays rapidly with further increasing Ir concentration so that the samples with $x$\,=\,0.75 and 1 are purely superconducting. By this choice we cover all phases of interest with two samples each for our systematic investigation of the orbital anisotropy.

\subsection{XAS data of CeRh$_{0.8}$Ir$_{0.2}$In$_5$} As an example, the top of Fig.\,\ref{Fig3} shows the total-electron-yield (TEY) intensities of the cerium $M_4$ and $M_5$ edge of CeRh$_{0.8}$Ir$_{0.2}$In$_5$ for the two polarizations E\,$\bot$\,$c$ (blue) and E\,$\parallel$\,$c$ (red) at T\,=\,8\,K. This temperature is low enough to assure excited CEF states are not populated ($\Delta E_1$\,$\approx$\,70\,K and $\Delta E_2$\,$\approx$\,230\,K) \cite{ChristiansonPRB70,ChristiansonPRB66,WillersPRB81}, so that the clear difference between the two polarizations is representative for the out-of-plane anisotropy of the ground state orbital. The green dots in the bottom panel of Fig.\,\ref{Fig3} present the resulting linear dichroism (LD) in a tenfold enlargement. The LD is the difference of the intensities for E\,$\bot$\,$c$ and E\,$\parallel$\,$c$. The data were then analyzed with an ionic full multiplet calculation (see section Analysis in Materials and Methods).

The CEF ground state in this compound family is a $\Gamma_7$ Kramer's doublet \cite{ChristiansonPRB70,ChristiansonPRB66,WillersPRB81} and can be expressed in terms of $J_z$ as
\[
\Gamma_7 = \alpha|\pm5/2\rangle + \sqrt{1-\alpha^2}|\mp 3/2
\rangle.
\]
The quantity $\alpha^2$ determines the out-of-plane anisotropy where $\alpha^2 > 1/6$ ($\alpha^2 < 1/6$) corresponds to a more
oblate (prolate) $4f$ orbital.  We find that a $\Gamma_7$ ground state with $\alpha^2$\,=\,0.387\,$\pm$\,0.005 describes the
data (see middle panel of Fig.\,\ref{Fig3}) and in particular the LD (see black line in the bottom panel) very well. The inset visualizes the corresponding $4f$ orbital. Note, dipole experiments like inelastic neutron scattering or soft XAS determine $\alpha^2$, so that the sign of $\alpha$ cannot be determined.

\subsection{Linear dichroism of all compositions} 
\begin{figure}
\centering
\includegraphics[width=.85\columnwidth]{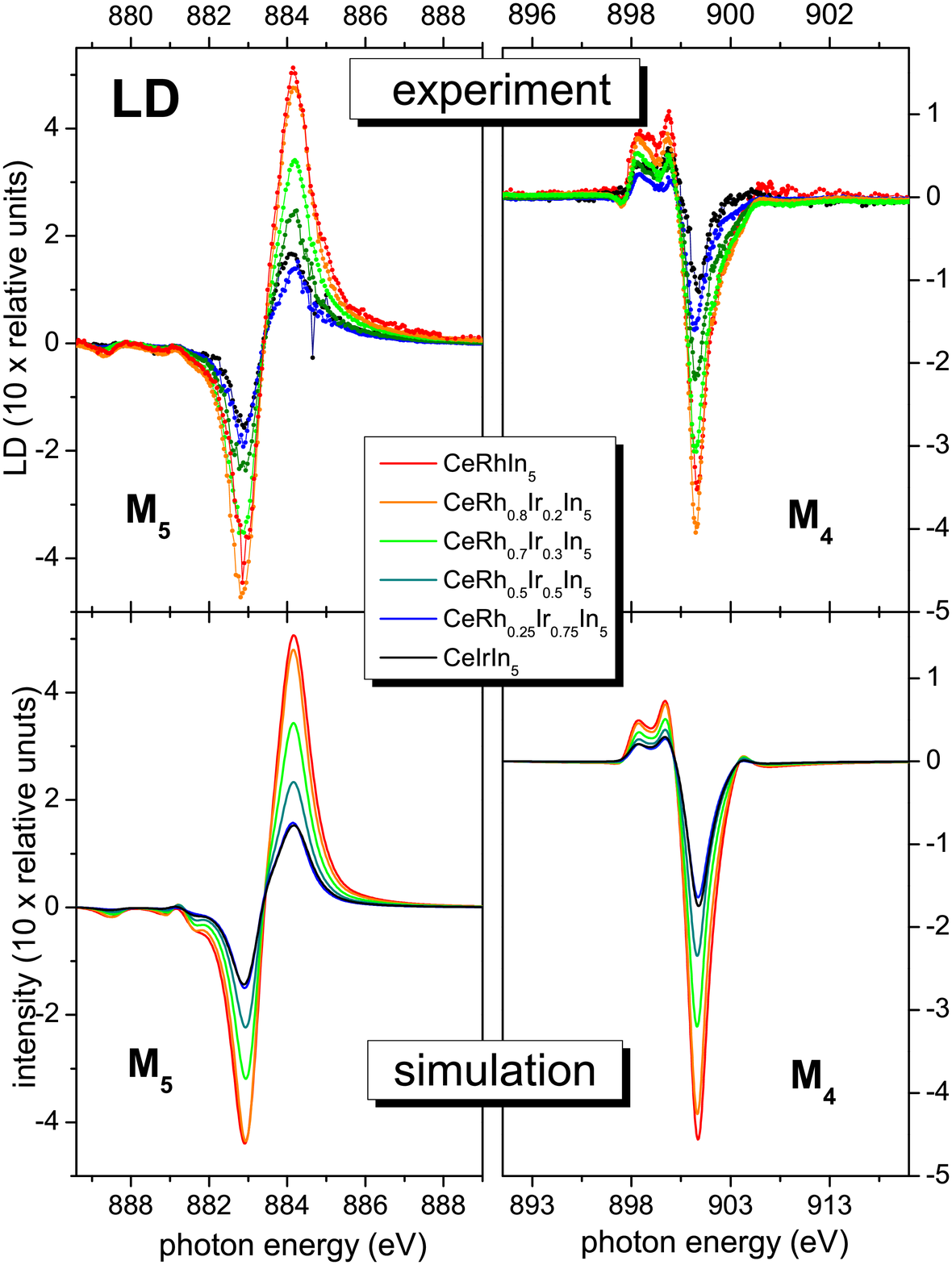}
\caption{The top panels show the experimental linear dichroism (LD) at the $M_5$ and $M_4$ edge for the Ir substitutions $x$\,=\,0.2, 0.3, 0.5, and 0.75 in CeRh$_{1-x}$Ir$_x$In$_5$ at 8\,K. Note, the data for $x$\,=\,0 and 1 were taken in a previous experiment \cite{WillersPRB81}. The bottom panels show the corresponding simulations which reproduce the data very well.}
\label{Fig4}
\end{figure}

The results for the other Ir concentrations, which were obtained in the same manner, are summarized in Fig.\,\ref{Fig4}. In the top panels of Fig.\,\ref{Fig4} the measured LD of both edges is depicted for the entire series of CeRh$_{1-x}$Ir$_{x}$In$_5$ from $x$\,=\,0 to 1. The LD is largest for the two Rh rich concentrations $x$\,=\,0 and 0.2 (red and orange dots). It then decreases rapidly for the intermediate Ir concentration $x$\,=\,0.3 (light green) and even more so for $x$\,=\,0.5 (dark green). The LD is smallest and almost the same on the Ir rich side ($x$\,=\,0.75 and 1, light and dark blue dots). The lower panels of Fig.\,\ref{Fig4} exhibit the corresponding simulated LD, which reproduce nicely the measured strong reduction of the LD with increasing Ir concentration. The resulting orbital anisotropies $\alpha^2$ are listed in Table\,1. It is important to note that the LD of the interim compositions cannot be simulated with the respective fractions of the LD of CeRhIn$_5$ and CeIrIn$_5$. This is most obvious for CeRh$_{0.5}$Ir$_{0.5}$In$_5$, where 0.5\:LD$_{\rm CeRhIn_5}$ + 0.5\:LD$_{\rm CeIrIn_5}$ would yield an $\alpha^2$ = 0.321 while a value of 0.284 has bee observed. This change of LD indicates that the CEF ground state wave function changes monotonically, but not linearly with the Ir concentration, unlike the lattice anisotropy in Fig.\,\ref{Fig2}b. For completeness we also give the value for CeCoIn$_5$ in Table\,1 \cite{WillersPRB81}.

\section{Discussion}
\subsection{Orbital anisoptroy $\alpha^2$ and phase diagram} 
\begin{figure*}
\begin{center}
\centerline{\includegraphics[width=0.7\textwidth]{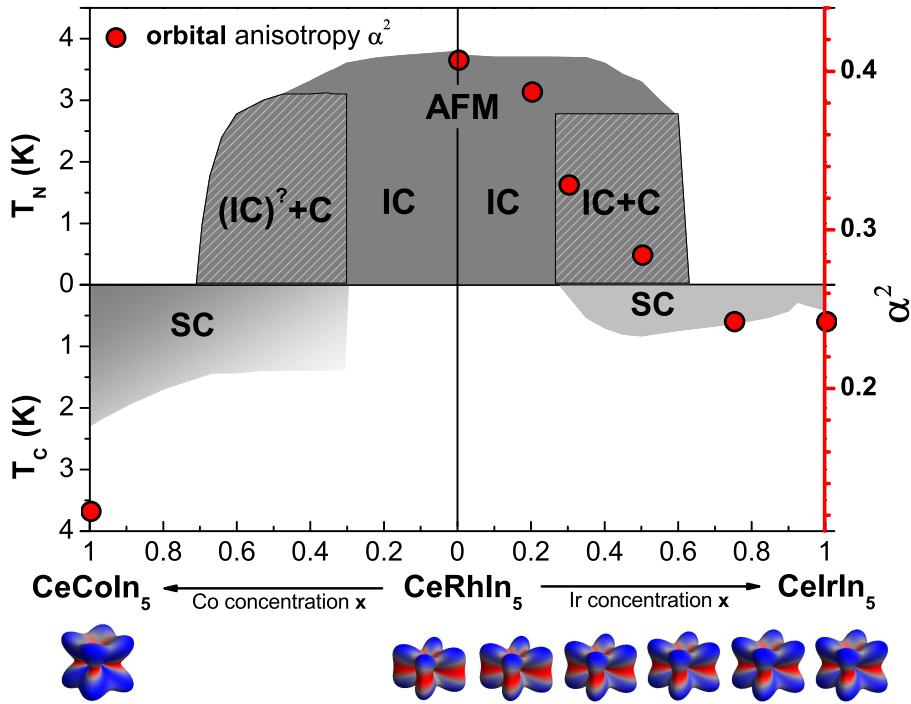}}
\caption{Orbital anisotropy $\alpha^2$ on the right scale as function of $x$ in CeRh$_{1-x}$Ir$_x$In$_5$ and for CeCoIn$_5$. The evolution with $x$ of the angular distributions of the $4f$ CEF ground-state orbitals in CeRh$_{1-x}$Ir$_x$In$_5$ and CeCoIn$_5$ is shown below the phase diagram.}
\label{Fig5}
\end{center}
\end{figure*}

In Fig.\,\ref{Fig5} the $\alpha^2$ values are plotted as red circles along with the CeRh$_{1-x}$Ir$_x$In$_5$ phase diagram for illustrating the changes of the wave function with the Ir concentration. The size of the circles corresponds to the error bars. As shown in Fig.\,\ref{Fig5}, we now can observe a clear trend between the value of $\alpha^2$ and the ground state properties of CeRh$_{1-x}$Ir$_x$In$_5$: The incommensurate antiferromagnetically ordered (IC AFM) samples CeRhIn$_5$ and CeRh$_{0.8}$Ir$_{0.2}$In$_5$ on the Rh rich side have almost the same large LD (see also Fig.\,\ref{Fig4}), yielding the largest $\alpha^2$-values within the series; the $\alpha^2$-value of the $x$\,=\,0.2 sample being only 5\% smaller than the one of the pure Rh compound. In the intermediate region where the three phases coexist, $\alpha^2$ decreases rapidly with $x$. On the Ir rich side of the phase diagram, in the purely superconducting region, $\alpha^2$ is small and levels off, i.e it is identical for CeRh$_{0.25}$Ir$_{0.75}$In$_5$ and CeIrIn$_5$.

\begin{table*}
\renewcommand{\arraystretch}{1.25}
\begin{tabular*}{\hsize}
{@{\extracolsep{\fill}}lccccccc}
        \hline
        \hline
					sample	&CeRhIn$_5$			&					&CeRh$_{1-x}$Ir$_x$In$_5$			&					&					&CeIrIn$_5$			& CeCoIn$_5$\\
					Ir concentration $x$		&  0   					&0.2			&0.3					&0.5			&0.75			&1							&						\\
        \hline
				$\alpha^2$	&0.407	&0.387	&0.328	&0.284	&0.242	&0.242	&0.123\\
				$\mu^{par}_{CEF}$ ($\mu_B$) &1.31	&1.30	&1.24	&1.23	&1.20	&1.20	&1.20\\
				Ce-In2 (\AA)	&3.2775	&-		&-		&-		&-		&3.272	&3.288\\
				\hline
        \hline
    \end{tabular*}
\caption{Out-of-plane anisotropy $\alpha^2$ ($\Delta$$\alpha^2$\!=\!0.005) for all measured Ir concentrations $x$ in CeRh$_{1-x}$Ir$_x$In$_5$ and for CeCoIn$_5$ from the present analysis. Note, $\alpha^2 = 1/6 \approx\ 0.166$ would correspond to a cubic-type orbital. Ce-In2 are the distances between cerium and out-of-plane indium (In2) as taken from Ref.\,\cite{Moshopulu_2002}. $\mu^{par}_{CEF}$ are the paramagnetic moments as calculated from the CEF ground states.}
\end{table*}

Obviously, the superconducting compositions favor the orbitals with smaller $\alpha^2$ values. This becomes even more evident 
when taking into account CeCoIn$_5$: The $\alpha^2$ value of CeCoIn$_5$ is smallest while it has the highest T$_c$ (Fig.\,\ref{Fig5}). Actually, its $\alpha^2$ value falls rather nicely onto the phase diagram when using the same scaling as for the CeRh$_{1-x}$Ir$_x$In$_5$ substitution series. This strongly supports our conjecture that $\alpha^2$ is a parameter which correlates with the magnetic as well as the superconducting phases, in contrast to $a/c$ which only serves the superconducting samples.

The implication of the changing \textit{J$_z$} contribution on the spatial distribution of the $4f$ electrons in the CEF ground state is obvious when recalling the pure $J_z$\,=\,$|\pm5/2\rangle$ is donut shaped, while the pure $|\pm3/2\rangle$ is yo-yo shaped \cite{HansmannPRL100}. Upon going from CeRhIn$_5$ to CeIrIn$_5$ the orbital extends increasingly out-of-plane (see Fig.\,\ref{Fig5}) such that it becomes less oblate with increasing Ir concentration. For CeCoIn$_5$ the $4f$ orbital is prolate, i.e. has the largest extension in the $c$ direction and CeCoIn$_5$ has the highest superconducting transition temperature. We discuss possible correlations and implications in the following.

Intuitively one would expect that the lattice anisotropy follows the orbital anisotropy, a behavior which has been observed e.g. in the Ce-monopnictides \cite{Hannan_2004}. However, in the Ce$M$In$_5$ family this is obviously not the case. While the lattice constants of the CeRh$_x$Ir$_{1-x}$In$_5$ series nicely follow Vegard's law \cite{Pagliuso_2001}, so that the \textit{lattice} anisotropy $a/c$ in Fig.\,\ref{Fig2}b changes accordingly with the Ir concentration, the \textit{orbital} anisotropy $\alpha^2$ changes in a strongly non-linear manner (see Fig.~\ref{Fig5}). Furthermore, while the unit cell becomes shorter along the $c$-axis from Rh to Ir, the $4f$ orbital extends increasingly out of the $ab$ plane, suggesting an anti-correlation, rather than a correlation, between the $4f$ wave function's \textit{flatness} and the $a/c$ ratio. However, CeCoIn$_5$ indicates the opposite (i.e. correlating) behavior: in this compound, the $4f$ orbital is the most extended one along $c$, while the smallest $a/c$ ratio (largest $c/a$) is observed. Apparently, there is no obvious systematics between orbital and lattice anisotropy (compare Fig.\,\ref{Fig2}b and Fig.\,\ref{Fig5}).

In Table\,1 we also list the calculated para\-magnetic moments that correspond to the CEF wave functions of CeRh$_x$Ir$_{1-x}$In$_5$ and CeCoIn$_5$. If we use the simple \textit{ansatz} that the strength of the exchange interaction $J_{ex}$ is proportional to the square of the paramagnetic moment, then we observe that $J_{ex}$ does not increase upon going from CeRhIn$_5$ to CeIrIn$_5$ or CeCoIn$_5$; instead, the trend is opposite: it decreases. Consequently, it is questionable whether it is an increase of $J_{ex}$ which changes the balance between the RKKY (T$_{RKKY} \propto N(E_F)J_{ex}^{2}$) and the Kondo ($T_K \propto \exp(-1/N(E_F)J_{ex})$) interaction strengths such that CeRhIn$_5$ is dominated by the RKKY and CeIrIn$_5$ and CeCoIn$_5$ by the Kondo interaction. There must be another mechanism or condition at play which changes the balance between the two interactions.

The evolution of anisotropy in ground state wave functions reflects the momentum dependence of the Ce $4f$ -- conduction band hybridization. In fact, this very aspect has been addressed by Shim \textit{et al.} \cite{Shim_2007}: their DMFT calculations for CeIrIn$_5$ show the Ce $4f$ moment to be more strongly coupled to the out-of-plane indium atoms (In2) than to the in-plane indium (In1) despite nearly identical Ce-In bond lengths (see Fig.\,\ref{structure}).

Following this line of thought for the other members of the Ce115 family, one would expect that CeCoIn$_5$ would show the least Ce-In(2) hybridization, and consequently should order magnetically since its Ce-In2 distances are largest. In contrast, CeRhIn$_5$ where the Ce and In2 are closer should be more itinerant with respect to CeCoIn$_5$ (see Table\,1) \cite{Moshopulu_2002}. This naive ansatz predicts a behavior that is opposite to the observed ground state properties of the Ce$M$In$_5$ compounds. This apparent contradiction is resolved by our experimental findings which reveal that the spatial distribution of the $4f$ wave function dictates the ground state: the $4f$ orbital of CeCoIn$_5$ has the largest extension in the $c$ direction, so that the large Ce-In2 distances are (over)compensated and an itinerant, nonmagnetic ground state forms. In CeRhIn$_5$, where the small $4f$ orbital extension in the $c$-direction leads to a reduction of the hybridization, the ground state is magnetically ordered. Our findings provide experimental evidence for the Shim \textit{et al.} conjecture that the out-of-plane hybridization is important.

How does this fit with the observed linear correlation between the lattice anisotropy $a/c$ and T$_c$ for the superconducting plutonium and cerium 115 compounds \cite{Bauer_2004}? We already noted that this only holds for the superconducting compositions, so that the $a/c$ ratio by itself cannot be used to predict whether a compound becomes superconducting. For example, the $a/c$ ratio of CeRhIn$_5$ would indicate a finite T$_c$ at ambient pressure, while in reality no superconductivity has been recorded (see Fig.\,\ref{Fig2}b). Here, we infer that $\alpha^2$ is the parameter that distinguishes superconducting from non-superconducting compounds. Once the purely superconducting region of the phase diagram has been reached, $\alpha^2$ does not change any more and the $a/c$ ratio seemingly tunes T$_c$.

Our observation shows that the orbital anisotropy is the driving parameter for the $cf$-hybridization, eventually resulting in superconductivity. We emphasize that the degree of hybridization has been measured by independent experiments. The stronger $cf$-hybridization of the superconducting compounds CeIrIn$_5$ and CeCoIn$_5$ is reflected in a larger Fermi surface \cite{Haga_2001,Fujimori_2003,Harrison_2004,Shishido_2005,Settai2007,Allen2013} and also in a larger quasielastic line width. For the latter, inelastic neutron scattering on powder samples found a line width of HWHM $\approx$ 1.4\,meV for the magnetically ordering Rh compound and line widths at least twice as large for CeIrIn$_5$ and CeCoIn$_5$ \cite{WillersPRB81}.

\subsection{Summary} In summary, the CEF ground state wave functions of the CeRh$_{1-x}$Ir$_x$In$_5$ substitution series have been determined and a clear correlation between ground state properties and wave functions has been observed. These findings suggest the $4f$ ground state orbital plays a decisive role in the detailed balance of RKKY and Kondo interactions and may explain why these seemingly similar materials have different ground states. More generally, anisotropic hybridization, in addition to $J_{ex}$, must be a necessary component in an appropriate description of Kondo lattice materials and of the evolution of their ground states as a function of a non-thermal tuning parameter.

\section{Materials and Methods}

\subsection{Samples} High-quality single crystals of CeRh$_{1-x}$Ir$_x$In$_5$ were grown by flux-growth and well characterized by magnetic
susceptibility and/or heat capacity to ensure their nominal composition is in accordance with the phase diagram \cite{Pagliuso_2001,Llobet_2005}. Before the absorption experiment all crystals were aligned within 1-2$^{\circ}$ by Laue x-ray diffraction. 

\subsection{Method} The linearly polarized soft XAS experiment was performed at the European Synchrotron Radiation Facility (ESRF) in Grenoble, France, at the ID08 undulator beamline. We recorded the Ce $M$-edge absorption spectra in the total electron yield mode
(TEY). The energy resolution at the $M_{4,5}$ edges ($h\nu \approx 875-910$\,eV) was set to 0.2\,eV. The samples were cleaved \textit{in situ} in an ultra high vacuum chamber with a pressure of 2\,x\,10$^{-10}$\,mbar at 8\,K. The $M_{4,5}$ edges were recorded for light polarized parallel (E\,$\parallel$\,$c$) and perpendicular (E\,$\bot$\,$c$) to the tetragonal $c$ axis. Here the undulator beamline has the advantage that for normal incidence the polarization can be changed without moving the sample, so that the same sample spot is probed for both polarization directions. For each sample different spots were probed in order to rule out sample inhomogeneities. The data of CeRh$_{1-x}$Ir$_x$In$_5$ were taken at 8\,K for the four Ir substitutions $x$\,=\,0.2, 0.3, 0.5, and
0.75. The excited CEF states are above 5\,meV ($\approx 50$\,K) \cite{ChristiansonPRB70,ChristiansonPRB66,WillersPRB81} so that at
8\,K only the ground state is probed. The data of CeRhIn$_5$ ($x$\,=\,0), CeIrIn$_5$ ($x$\,=\,1), and also of CeCoIn$_5$ were taken in
a previous experiment \cite{WillersPRB81}. For the data analysis all data were normalized to the intensity of the isotropic spectra I$_{iso}$\,=\,(2I$_{E\bot c}$~+~I$_{E||c}$)/3.

\subsection{Analysis} The XAS data have been simulated with ionic full multiplet calculations based on the XTLS 8.3 program \cite{TanakaJPSC63}. The atomic parameters are given by reduced Hartree-Fock values. The reduction accounts for the configuration interaction which is not included in the Hartree-Fock calculations and is determined from fitting the isotropic spectra I$_{iso}$. Typical reductions are about 40\% for the $4f-4f$ Coulomb interactions and about 20\% for the $3d-4f$ interactions \cite{HansmannPRL100,WillersPRB81,StrigariPRB86}. In the present manuscript the LD of the end members of the series were analyzed in the same manner as for the substitution series investigated here.

\begin{acknowledgments}
We thank P.~Thalmeier for fruitful discussions. The wave function density plots were calculated using the CrystalFieldTheory package for Mathematica written by M.~W.~Haverkort. Support from the German funding agency DFG under Grant No. 583872 is gratefully acknowledged. Work at LANL was performed under the auspices of the U.S. DOE, Office of Basic Energy Sciences, Division of Materials Sciences and Engineering.
\end{acknowledgments}

\end{document}